# Strain-induced spin states in atomically-ordered cobaltites

*Woo Seek Choi*[1], *Ji-Hwan Kwon*[2], *Hyoungjeen Jeen*[1], *Jorge E. Hamann-Borrero*[3,4], *Abdullah Radi*[3], *Sebastian Macke*[5,6], *Ronny Sutarto*[7], *Feizhou He*[7], *George A. Sawatzky*[8], *Vladimir Hinkov*[5], *Miyoung Kim*[2], *and Ho Nyung Lee*[1*]

[1]Materials Science and Technology Division, Oak Ridge National Laboratory, Oak Ridge, TN 37831, USA

[2]Department of Materials Science and Engineering, Seoul National University, Seoul, 151-744, Korea

[3]Quantum Matter Institute, University of British Columbia, Vancouver, V6T1Z1, Canada

[4]Leibniz Institute for Solid State and Materials Research, IFW Dresden, 01171 Dresden, Germany

[5]Max Planck-UBC Centre for Quantum Materials, Vancouver, V6T1Z1, Canada

[6]Max Planck Institute for Solid State Research, Heisenbergstr. 1, 70569 Stuttgart, Germany

[7]Canadian Light Source, University of Saskatchewan, Saskatoon, Saskatchewan, S7N0X4, Canada

[8]Department of Physics and Astronomy, University of British Columbia, Vancouver, V6T1Z1, Canada

**Epitaxial strain imposed in complex oxide thin films by heteroepitaxy is recognized as a powerful tool for identifying new properties and exploring the vast potential of materials performance. A particular example is $LaCoO_3$, a zero spin, non-magnetic material in the bulk, whose strong ferromagnetism in a thin film remains enigmatic despite a decade of intense research. Here, we use scanning transmission electron microscopy complemented by x-ray and optical spectroscopy to study $LaCoO_3$ epitaxial thin films under different strain states. We observed an unconventional strain relaxation behavior resulting in stripe-like, lattice modulated patterns, which did not involve uncontrolled misfit dislocations or other defects. The modulation entails the formation of ferromagnetically ordered sheets comprising intermediate or high spin $Co^{3+}$, thus offering an unambiguous description for the exotic magnetism found in epitaxially-strained $LaCoO_3$ films. This observation provides a novel route to tailoring the electronic and magnetic properties of functional oxide heterostructures.**

Epitaxial strain induced in a thin film due to lattice mismatch with a substrate serves as a versatile controlling parameter of physical and chemical properties, including the electronic structure, conductivity, ionic diffusion, magnetism, ferroelectricity, and crystallographic symmetry. Not only can it fine tune the properties, but sometimes stabilize novel behaviors, which cannot be found in bulk counterparts. Heteroepitaxial perovskite oxides are a fascinating model system to explore the strain physics, as they offer prototypical $ABO_3$ building blocks that are structurally alike, while exhibiting a wide spectrum of useful physical properties. For example, recent studies showed that scientifically and technologically important properties such as magnetoresistance, ferroelectricity, multiferroicity, and flexoelectricity could be effectively controlled, stabilized or enhanced by employing epitaxial strain in a perovskite heterostructure.[1-4] In epitaxial thin films, the mechanical stress due to the lattice mismatch between the film and substrate increases with the film thickness. Above the critical thickness ($h_c$), the film cannot sustain the substrate-imposed lattice spacing anymore. Conventionally, in such cases, misfit dislocations are formed and the film gradually relaxes by adopting its bulk lattice structure with increasing thickness. The dislocations are distributed irregularly and constitute a poorly controlled relaxation mechanism that can have a catastrophic impact on device performance.

Lanthanum cobaltite ($LaCoO_3$, LCO) is a very interesting material in this context, as we have observed a completely new strain-relaxation behavior when grown in epitaxial film forms. LCO has a stable $Co^{3+}$ valence state with $3d^6$ orbital configuration and rhombohedral crystal structure ($a$ = 5.3781 Å, $\beta$ = 60.81° and pseudocubic $a_{pc}$ = 3.8029 Å).[12] More interestingly, thin film LCO has been reported to exhibit a long range ferromagnetic ordering below ~80 K, whereas bulk LCO shows a non-magnetic ground state. Previous studies commonly observed this magnetic ordering,[5-11] and some of the studies attributed the magnetic ordering to the epitaxial strain.[5-8] However, the exact origin of the exotic magnetic properties of LCO films has not yet been clearly unveiled. It is also known that LCO is a good ferroelastic material showing spontaneous strain,[13-15] which makes it even more intriguing to study the strong strain coupling of the structural, electronic, ionic, and magnetic properties. In addition, cobalt-based perovskite oxides are promising candidates for ionic conductors and surface catalysts, and the strain can be used as a tool to control the ionic activities as predicted by density functional theory calculations.[16,17] Therefore, the strain-dependent behaviors can bring new insight for such technological applications as well.

To understand the microscopic origin of the ferromagnetism and the effect of strain, we have grown high quality epitaxial LCO thin films on various substrates: $LaAlO_3$ (LAO), $(LaAlO_3)_{0.29}(SrAl_{0.5}Ta_{0.5}O_3)_{0.71}$ (LSAT), and $SrTiO_3$ (STO) to impose different strain states. We systematically characterized the films with various techniques including Z-contrast scanning transmission electron microscopy (STEM), x-ray absorption (XAS), and ellipsometric spectroscopy, complemented by magnetometry and x-ray diffraction (XRD).

XRD results (Figure 1) confirmed the excellent epitaxial growth of our films by pulsed laser epitaxy. Only LCO film and substrate peaks were observed in XRD $\theta$-$2\theta$ scans, without any secondary phases (Figure 1a). A systematic decrease in the $c$-axis lattice constant of LCO was observed as the substrate lattice constants increase from LAO (compressive strain $\varepsilon = -0.3\%$), LSAT (tensile strain $\varepsilon = 1.7\%$) to STO (tensile strain $\varepsilon = 2.6\%$), as shown in Figure 1b. Clear Kiessig fringes also assured a good structural quality of the thin films. Figures 1c-1e show XRD reciprocal space maps (RSM) around the substrates' 114 Bragg reflections. In addition to the systematic change in the *c*-axis lattice constant, all LCO films were coherently grown on the substrates. This indicates that the in-plane lattice constants of all three samples are coherently maintained as their substrate, at least up to a thickness of 30 nm.

Figures 2a, 2b, and 2c show STEM images for LCO films grown on LAO, LSAT, and STO substrates, respectively. Besides testifying to the high quality epitaxial growth with atomically sharp interfaces and surfaces, the images exhibit systematic strain-dependent distinct dark stripes. For the LCO film grown under slight compressive strain on LAO, only a few in-plane stripes were rarely observed (Figure 2a). As tensile strain was applied, on the other hand, LCO started to show stripes running perpendicularly to the surface (Figure 2b). As the degree of tensile strain increased, more such stripes appeared, eventually forming a fairly regular superstructure ($3a_0$ periodicity) as shown for the LCO film on STO (Figure 2c). Note that the lattice planes of film and substrate are perfectly aligned without developing any misfit dislocations, which would be readily imaged by STEM. We also note that there were a few parts in the film with actual dislocations, but in those parts, the dark stripes were not observed.

The most probable origin and natural explanation of the dark stripes is a modulation of the atomic arrangement, *i.e.* lattice modulation. Previously, oxygen vacancies and their orderings have been considered as the cause of lattice expansion and resultant dark stripes in TEM, particularly in Sr-doped and oxygen deficient $LaCoO_{3-\delta}$ films.[18-22] However, if oxygen vacancies were responsible for the stripes in our case as well, their high density, in particular for the film on STO, would imply an oxygen deficiency $\delta$ of well above 0.1 which should be easily detectable by XAS and/or spectroscopic ellipsometry. In Figure 3a, we show XAS around the Co *L*-edges for the LCO film on STO measured in the bulk sensitive fluorescence yield (FY) mode. By comparing the spectral shape with that from the stoichiometric bulk LCO,[23] we could confirm a good stoichiometry of our thin film. In particular, $Co^{2+}$ would inevitably accompany oxygen deficiency, resulting in typical fingerprints like a distinctive prepeak just before the Co $L_3$ edge.[24] Note that most of the previous reports of XAS on epitaxial LCO films showed this prepeak, in contrast to our result.[9,25,26] From the XAS data, we could put an upper limit of the $Co^{2+}$ proportion to the total signal of less than 5%. In addition, usually, the electronic structure changes drastically by even a small compositional change or doping in transition metal oxides. The resulting change can be readily probed by optical spectroscopy, as it is an excellent tool to characterize such a stoichiometry change both qualitatively and quantitatively.[27,28] Figure 3b shows the spectroscopic ellipsometry result for the LCO films. Optical conductivity, $\sigma_1(\omega)$ of bulk LCO is also shown for comparison. If the LCO films were off-stoichiometric even by a small amount (< 5%), it should result in a drastic change in $\sigma_1(\omega)$.[27,28] More importantly, $\sigma_1(\omega)$ in the LCO films should also reveal systematic evolution as the substrate's lattice constant is gradually increased, if the stripe formation were due to oxygen vacancies. However, as shown in Figure 3b, no significant change was observed in $\sigma_1(\omega)$ of the films, indicating that our LCO films had a good stoichiometry without containing a discernible amount of oxygen vacancies. Based on our two different spectroscopic analyses, we conclude that the oxygen vacancy is not the main cause of the stripe pattern in our LCO films.

Before further discussing the origin and implications of the lattice modulation, we note that the systematic strain dependence of the stripe density suggests a unique mechanism for strain relaxation. Considering conventional strain relaxation, $h_c$ of LCO films on STO and LSAT are about 1.8 and 3.4 nm, respectively, which roughly corresponds to four and eight unit cells.[29] The XRD RSM results, however, excluded any notable macroscopic strain relaxation even though the films were much thicker than the calculated $h_c$. This suggests that a certain type of local structural change should be made to compensate the huge tensile strain. A closer examination of the STEM images showed two distinct regions in the LCO films under tensile strain (Figures 2d and 2e). I: The structurally uniform region (< ~$h_c$) close to the interface with the substrate, free from any vertical dark stripes. II: The region (> $h_c$) with distinct vertical dark stripes. Within this scenario, the near absence of the stripes in the LCO film on LAO is naturally explained by the fact that the corresponding $h_c$ of about 26 nm exceeds the film thickness.

Such an unconventional strain relaxation mechanism has not been observed in other materials, and we attribute it to the strong ferroelasticity in LCO.[13-15] A closer inspection of the STEM image in Figure 2c revealed that the crystal lattice symmetry was only lowered but not destroyed: Both regions are still strained to the substrate – the region I in the conventional sense (*i.e.* both in XRD and STEM) and the region II only on macroscopic scale (only in XRD). A configuration with conventional strain relaxation with globally contracted in-plane lattice constants would follow only for a much thicker film,[11] and would almost always involve misfit dislocations or other structural defects.

Now, in order to assess the nature of the lattice modulation-induced stripe patterns, we mapped out inter-atomic distances from the STEM image shown in Figure 2c. Figure 4a shows a part of the experimental STEM image for the LCO film on STO. The intensity profiles for the red and green lines drawn in Figure 4a, and the inter-atomic spacing between the neighboring La atoms are shown in Figures 4b and 4c, respectively. While region I (green) showed more or less equal spacing between the La atoms, region II (red) showed clear spatial modulation, which resulted in vertical dark stripes in the image. Based on the atomic spacing extracted from the experimental image, we simulated a LCO film structure, as shown in Figures 4e (cross sectional view) and 4f (top view). For region I, since this region was fully strained and had a uniform atomic distance, we applied a regular tetragonal distortion. For region II, in order to accommodate the observed modulation and the resulting lower symmetry, we imposed a nano-scale monoclinic distortion. In particular, to incorporate the $3a_0$ modulation, we constructed alternating two monoclinic and one tetragonal unit cells along the [010] direction. We further assumed the monoclinic domains to have an opposite angular distortion to the neighboring domains, as indicated by the yellow and orange lines in Figure 4f (top view). Interestingly, the resulting STEM image simulation for the artificially-constructed LCO structure conspicuously reproduced the experimental image (Figure 4d), supporting the explanation of the dark stripes in terms of a monoclinic nano-domain formation. A similar structural $3a_0$ modulation due to atomic-scale ferroelastic twinning was also previously observed in a mechanically-stressed, polycrystalline stoichiometric LCO,[14,15] strongly supporting our conclusion. We note that future local spectroscopic studies, *e.g.* electron energy loss spectroscopy, might provide more detailed information on the stripe patterns.

The observed microscopic structural modulation was strongly coupled to the unusual macroscopic ferromagnetic ordering in the LCO films. Figure 5 shows temperature dependent magnetization, $M(T)$ and magnetic field dependent magnetization, $M(H)$ curves of the LCO films with $H$ applied along the in-plane direction. In $M(T)$ curves (Figure 5a), we observed a similar ferromagnetic transition with $T_C$ = ~80 K for both LCO films on LSAT and STO, as reported in the literature.[5-10] $M(H)$ curves in Figure 5b also show clear ferromagnetic hysteresis loops with similar coercive fields. The saturation magnetization was somewhat larger for the LCO film on STO by a factor of ~1.15, compared to the film on LSAT. On the other hand, LCO on LAO did not show any discernible magnetic transition or hysteresis loop. Therefore, we claim that the epitaxial strain and the resultant atomic displacements promote the ordered spin state and long-range ferromagnetic ordering that is not stable in bulk. (Note that the horizontal stripes in the LCO film on LAO are much more scarce compared to the vertical stripes in the film on other substrates. See Supporting Information.)

It is known that an increase of the overall Co-O bond length in an octahedral coordination of only 0.1 Å can change the spin state from low to high spin.[30] In the present case, the Co-O bond length in the tetragonally distorted octahedra was increased by as much as 0.5 Å due to the local lattice distortion (Figure 4c), which could result in an electronic structure closer to that of a square planar coordination of $Co^{3+}$. Such an unusual coordination would give rise to a distinctively different orbital occupation and spin state compared to the previously discussed ones the literature. Depending on local details, the $Co^{3+}$ ion would have either intermediate spin ($S$ = 1) or high spin ($S$ = 2) state. This would provide large enough local magnetic moments to establish a long-range ferromagnetic ordering in the array of sheets separated by three lattice parameters indicated by the STEM result. In other words, we could expect that the tetragonally distorted Co octahedra have non-zero spin, while the monoclinically distorted Co octahedra with lower symmetry have zero spin. We can further estimate the number of tetragonally distorted octahedra within our 8 nm-thick films, which are about 35.7 and 40.0% for LCO films on LSAT and STO, respectively. The ratio of the population of tetragonally distorted octahedra between the LCO films on LSAT and STO is about 1.12, which is very close to the ratio in saturation magnetization between the films. Indeed, our experimental finding on ferromagnetism driven by crystallographic symmetry change is highly compatible to a theoretical prediction reported elsewhere,[31] which suggested that the suppression of octahedral rotations in combination with heteroepitaxial symmetry constraints (tetragonal distortion) could stabilize the ferromagnetic ground state. Instead of simple and global in-plane lattice expansion, however, our observation points out that the outcome of the tensile strain in ferroelastic LCO is far more sophisticated with local lattice modulation, which should be considered in future study of the material. We also note that mixed spin states have been recently suggested both experimentally and theoretically,[25,26,32] which further support our interpretation and the direct correlation between the local crystal structure and spin state.

In conclusion, we observed the strain-dependent microscopic lattice modulation in LCO epitaxial thin films, which yielded a novel spin state ordering. This observation highlights the collective phenomena in a transition metal oxide, whose physical property is globally modified by a local structural change. Moreover, the new mechanism responsible for the ferromagnetism can open a door to a better understanding of physical properties of cobalt-based oxides and to identifying new opportunities for many technologically relevant applications.

FIGURES

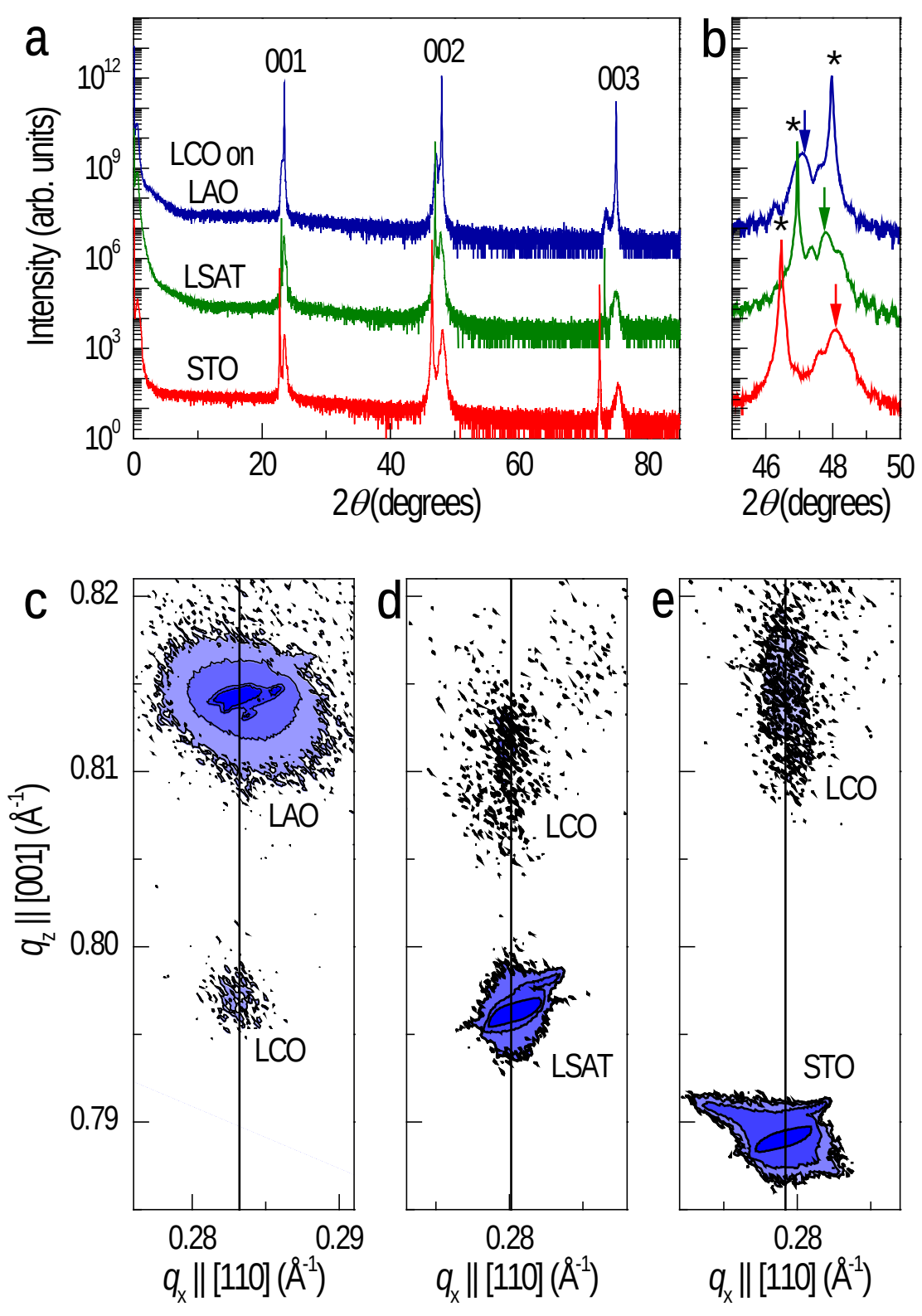

**Figure 1.** X-ray diffraction data of the studied LCO films. (a) X-ray $\theta$-2$\theta$ diffraction patterns of the films grown on STO (red), LSAT (green), and LAO (blue) substrates. (b) X-ray $\theta$-2$\theta$ diffraction patterns around the substrate 002 peaks. (The 002 Bragg reflection from the LCO films and the substrates are denoted as arrows and asterisks, respectively.) The *c*-axis lattice constants of the films show a systematic decrease with increasing the degree of tensile strain from LAO to STO. X-ray RSMs are shown around the 114 Bragg peak from (c) LAO, (d) LSAT, and (e) STO substrate.

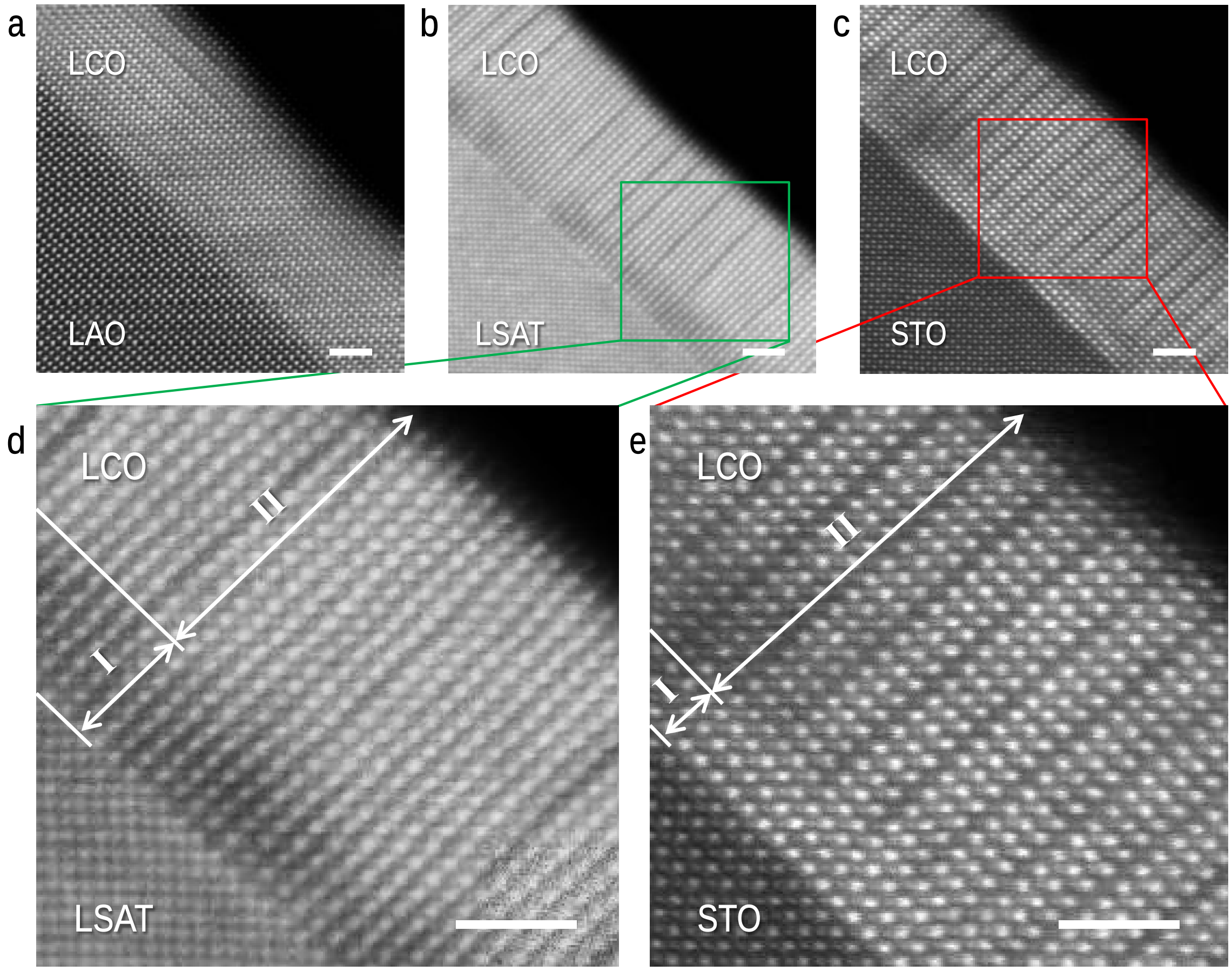

**Figure 2.** Microstructure and strain dependent lattice modulations. *Z*-contrast STEM images for LCO films grown on (a) LAO, (b) LSAT, and (c) STO. For LCO grown on LAO, only a few horizontal dark stripes are observed as in (a). As the tensile strain is applied, vertical dark stripes start to appear as in (b). When more tensile strain is applied, the vertical dark stripes become regular with a $3a_0$ periodicity as in (c). Two distinct regions observed in LCO films grown on (d) LSAT and (e) STO. Region I corresponds to the “uniform region” without any dark stripes. Region II corresponds to the “atomic ordering region” with the dark stripes. The LCO film on STO has a thinner uniform region I as compared to the film on LSAT due to the larger mismatch. The scale bars correspond to 2 nm.

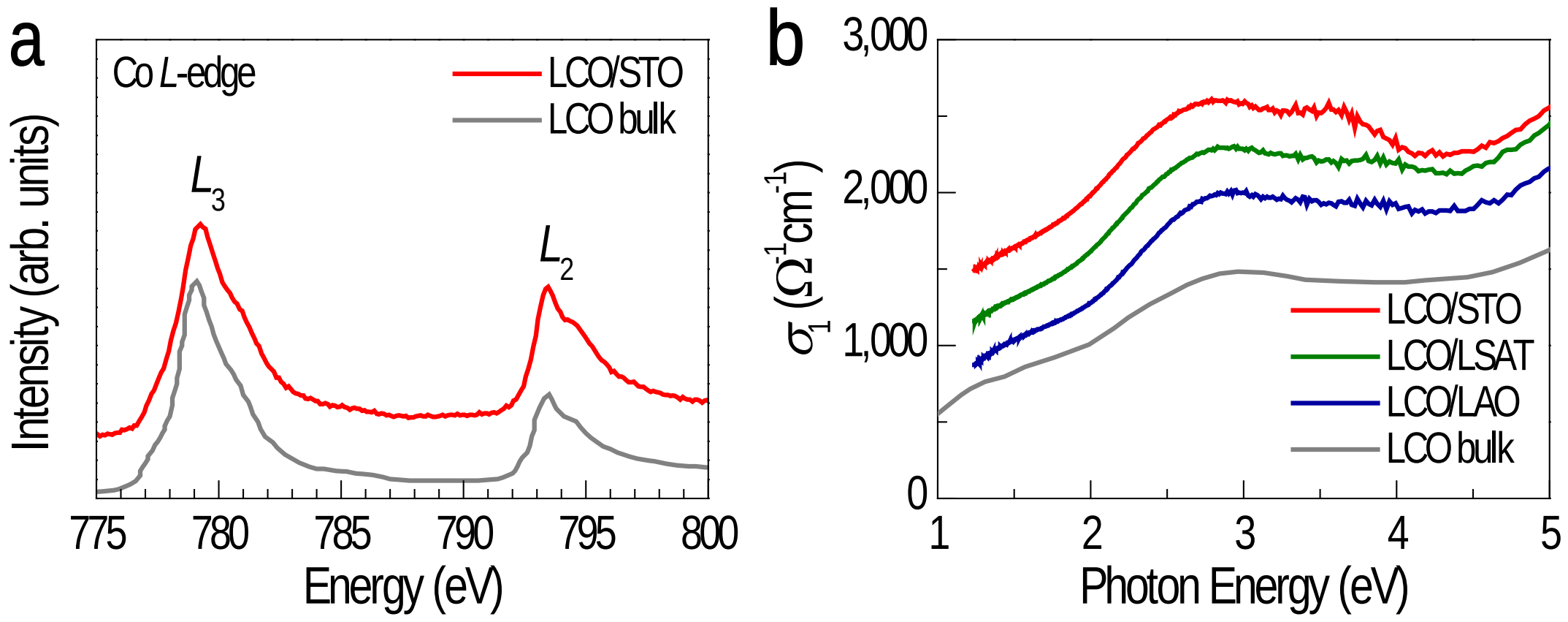

**Figure 3.** Strain-independent, robust electronic structure. (a) Co *L*-edge XAS for the LCO film on STO obtained in bulk sensitive FY mode. (b) $\sigma_1(\omega)$ for LCO grown on STO (red), LSAT (green), and LAO (blue). $\sigma_1(\omega)$ for bulk LCO is also shown as comparison. Each spectrum is shifted by 300 $\Omega^{-1}cm^{-1}$ from the last for clarity. Bulk XAS and $\sigma_1(\omega)$ are also shown in grey lines for comparison (refs. 23, 28).

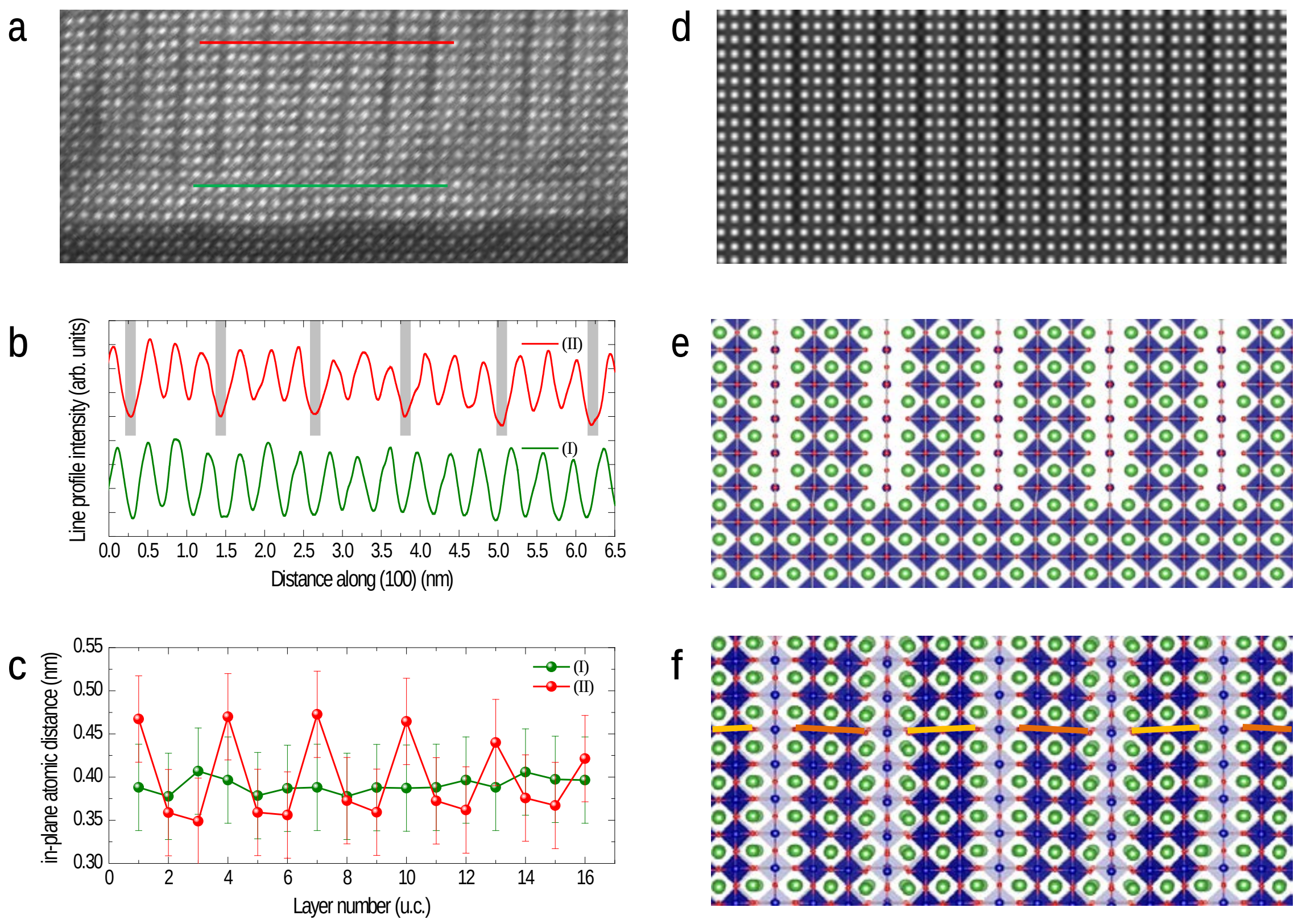

**Figure 4.** Image simulation of structural nano-domains. (a) Experimental, cross-sectional Z-contrast STEM image for the LCO film on STO (image taken from Figure 2(c)). (b) Intensity profile for the lines (green for region I and red for region II) drawn in (a) Grey areas indicate the dark vertical stripes for region II. (c) In-plane atomic distances between La atoms extracted from (b). (d) Simulated, cross-sectional STEM image for a LCO film on STO, clearly resembling the experimental image shown in (a). (e) Cross sectional and (f) top view for the artificially constructed LCO film structure. Yellow and orange lines in the top view image indicate the opposite monoclinic distortion for the neighboring domains.

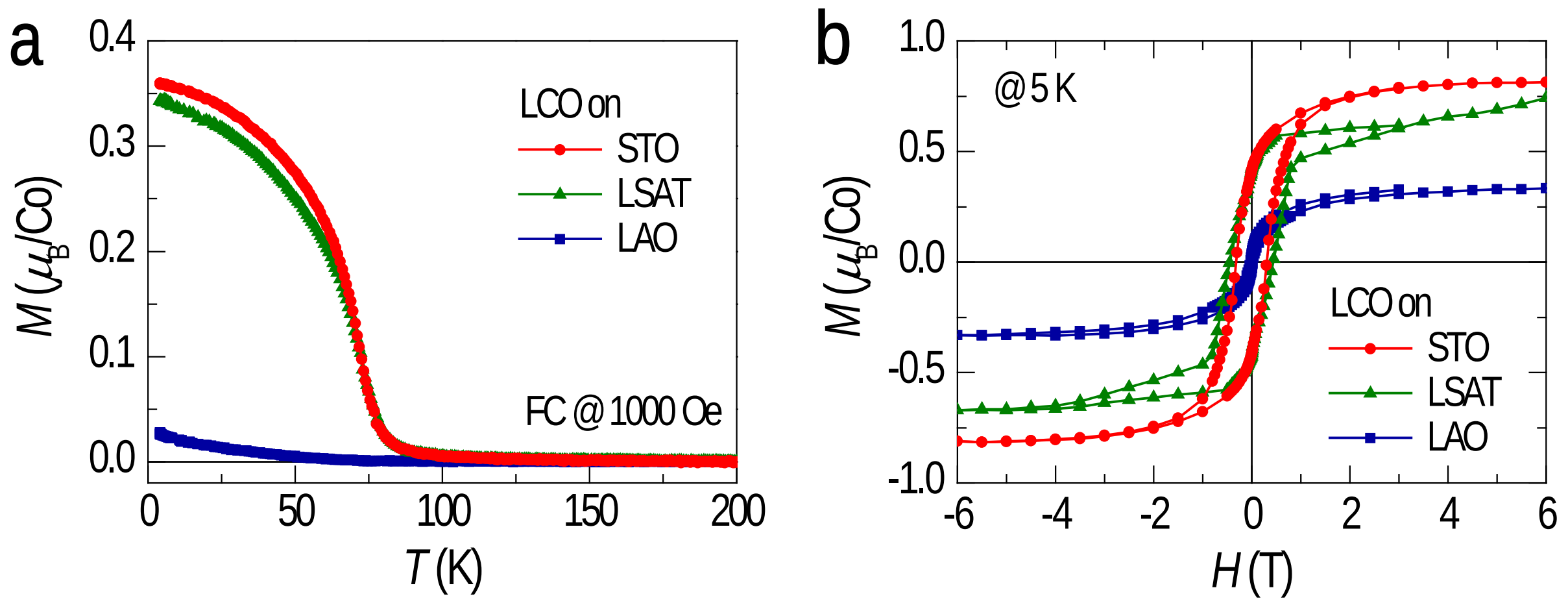

**Figure 5.** Strain-dependent magnetic properties. (a) *M*(*T*) and (b) *M*(*H*) curves for LCO on STO (red), LSAT (green), and LAO (blue). While the LCO film grown on LAO does not show much change, a clear ferromagnetic transition is apparent for the LCO films grown on LSAT and STO at ~80 K.

ACKNOWLEDGMENT

We appreciate valuable discussion with J. Lee, Y. M. Kim, M. W. Haverkort, H. Seo, and P. E. Vullum. We also thank J. Lee and Z. Gai for helping with the QSTEM simulation and magnetization measurement, respectively. This work was supported by the U.S. Department of Energy, Basic Energy Sciences, Materials Sciences and Engineering Division. The optical measurement was in part conducted at the Center for Nanophase Materials Sciences, a DOE-BES user facility. The research at the CLS was supported by NSERC, NRC, CIHR and the University of Saskatchewan. We acknowledge financial support from the Deutsche Forschungsgemeinschaft within the framework of the TRR80, project C1. The work at SNU was supported by the National Research Foundation of Korea grant funded by the Korea government (MEST) (No. 2012-0005637).

# Strain-induced spin-states in atomically-ordered cobaltites

*Woo Seok Choi*[1], *Ji-Hwan Kwon*[2], *Hyoungjeen Jeen*[1], *Jorge E. Hamann-Borrero*[3,4], *Abdullah Radi*[3], *Sebastian Macke*[5,6], *Ronny Sutarto*[7], *Feizhou He*[7], *George A. Sawatzky*[8], *Vladimir Hinkov*[5], *Miyoung Kim*[2], *and Ho Nyung Lee*[1*]

[1]Materials Science and Technology Division, Oak Ridge National Laboratory, Oak Ridge, TN 37831, USA

[2]Department of Materials Science and Engineering, Seoul National University, Seoul, 151-744, Korea

[3]Quantum Matter Institute, University of British Columbia, Vancouver, V6T1Z1, Canada

[4]Leibniz Institute for Solid State and Materials Research, IFW Dresden, 01171 Dresden, Germany

[5]Max Planck-UBC Centre for Quantum Materials, Vancouver, V6T1Z1, Canada

[6]Max Planck Institute for Solid State Research, Heisenbergstr. 1, 70569 Stuttgart, Germany

[7]Canadian Light Source, University of Saskatchewan, Saskatoon, Saskatchewan, S7N0X4, Canada

[8]Department of Physics and Astronomy, University of British Columbia, Vancouver, V6T1Z1, Canada

## THIN FILM GROWTH

We used pulsed laser epitaxy to grow LCO thin films on the three different substrates LAO, LSAT and STO. A KrF excimer laser ($\lambda$ = 248 nm) with a laser fluence of ~1 J/cm$^2$ was used for ablating a sintered LCO target. Thin film samples (8 – 30 nm in thickness) were fabricated at 700 °C in 100 mTorr of oxygen. Although here we do not detail the growth optimization, we note that the oxygen growth pressure is highly critical for synthesizing phase pure LCO thin films. We found that thin films grown at a lower pressure often resulted in unwanted cobalt oxide impurity and/or polycrystalline phases evidenced by x-ray diffraction (XRD).

## EXPERIMENTS

The crystal structure of LCO films was examined using XRD (X'Pert, Panalytical Inc.). The microstructure was investigated by aberration-corrected STEM (JEOL2200Fs) equipped with a CEOS 3$^{rd}$ order probe Cs corrector with a resolution of ~1 Å. The cutoff angle was 100 mrad. To simulate the STEM image, we used QSTEM (http://www.christophtkoch.com/stem/index.html). XAS experiments were carried out at the RSXS endstation of the REIXS beamline of the Canadian Light Source in Saskatoon with $\sigma$-polarized light. Spectroscopic ellipsometry (M-2000, J. A. Woollam Co.) was used to obtain $\sigma_1(\omega)$ between 1.24 and 5 eV at room temperature. $M(T)$ and $M(H)$ were recorded using a superconducting quantum interference device magnetometer (Quantum Design Inc.).

## DISAPPEARANCE OF MAGNETISM IN LCO FILM ON LAO SUBSTRATE

Figure S1 shows low resolution STEM image for LCO on LAO (left) and LSAT (right). The density of horizontal stripes in LCO on LAO is negligible compared to the vertical stripes found in LCO/LSAT and LCO/STO. This seems to be because the lattice parameter of LAO substrate is much closer to LCO compared to other substrates. In any case, the reduced number of stripes in LCO on LAO explains why the ferromagnetic signal is suppressed in the film.

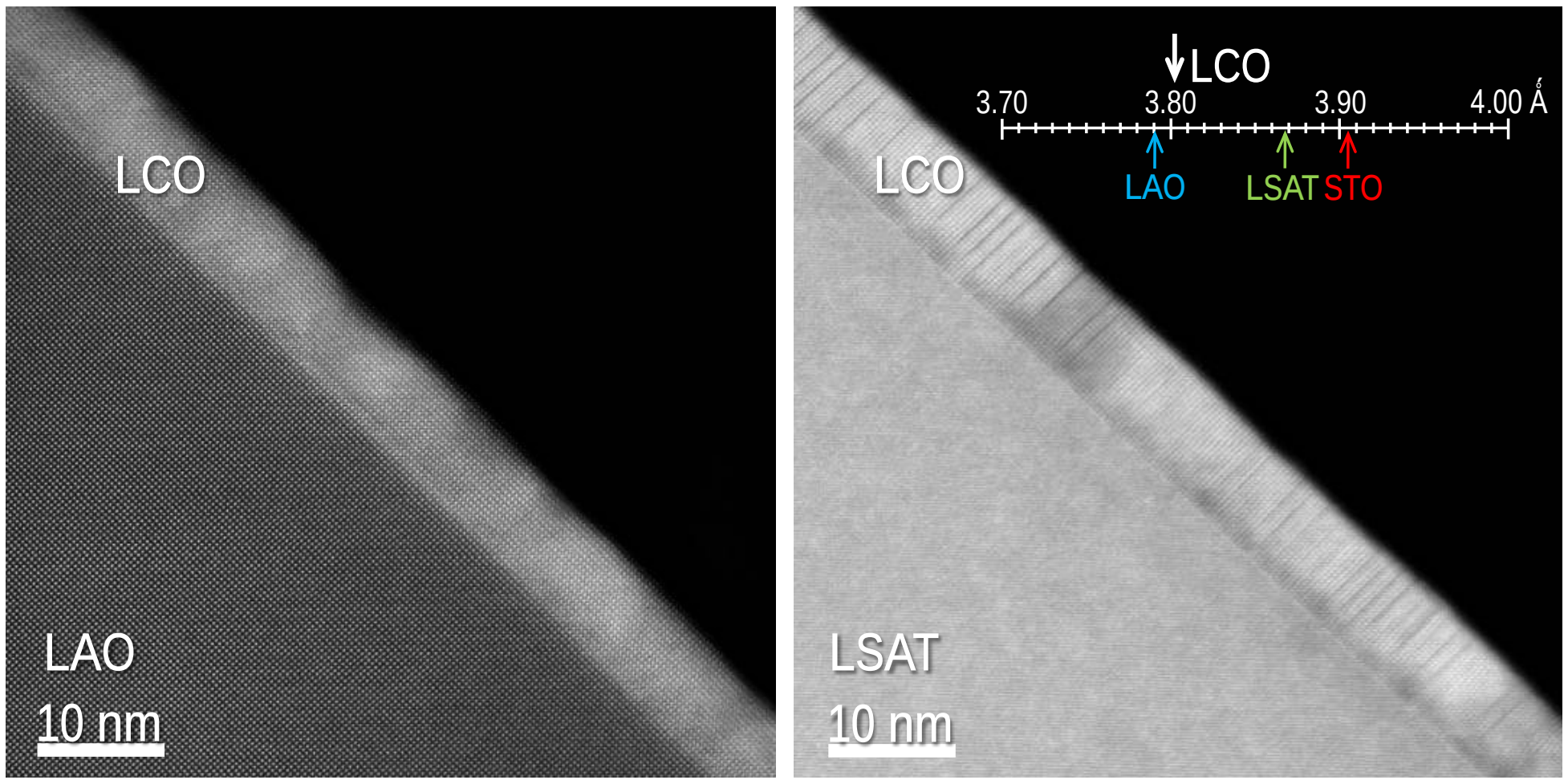

Figure S1. Low resolution STEM image for LCO on LAO and LSAT.

## EXCLUDING TEM SPECIMEN PREPARATION PROCEDURE AS A CAUSE OF THE STRIPE PATTERN

One might argue that exotic contributions, *e.g.* ionization damage during imaging or milling damage during TEM specimen preparation, could not be simply excluded as the origin of the dark stripes in the LCO films. Such extrinsic cause, however, is unlikely to play a major role due to the fact that the stripe pattern changes systematically with the strain state.